\documentclass[journal]{IEEEtran}
\IEEEoverridecommandlockouts
\usepackage{mathrsfs}
\usepackage{amsfonts}
\usepackage{ifpdf}
\usepackage{cite}

\ifCLASSINFOpdf
\usepackage[pdftex]{graphicx}
\else \fi
\usepackage[cmex10]{amsmath}
\usepackage{array}
\usepackage{mdwmath}
\usepackage{mdwtab}
\usepackage{eqparbox}
\usepackage[tight,footnotesize]{subfigure}
\usepackage{algorithmic}
\usepackage{algorithm}
\usepackage{amsmath}
\usepackage{epsfig}
\usepackage{color}
\usepackage{ae}
\usepackage{amssymb}
\usepackage{times}
\usepackage{amsmath}   % add

\usepackage{xcolor}

\usepackage{amsthm} % Let Lemma, Theorem, and Proposition be italic.

\usepackage{supertabular} % add
\usepackage{stfloats}
\usepackage{multirow}%用于表格
\newlength\savewidth
\newcommand\shline{\noalign{\global\savewidth\arrayrulewidth
                            \global\arrayrulewidth 1.5pt}%
                   \hline
                   \noalign{\global\arrayrulewidth\savewidth}}

\newtheorem{theorem}{\textbf{Theorem}}
\newtheorem{lemma}{\textbf{Lemma}}

\newtheorem{remark}{\textbf{Remark}}

\begin{document}
%----------------------------------------Make Title -----------------------------------------

%-------------------------------------------------------------------------------------------
%\title{Joint Mode Selection, Admission Control, Partner Assignment, and Power Allocation in Underlay SCMA Device-to-Device Networks}
\title{Underwater Anchor-AUV Localization Geometries with an Isogradient Sound Speed Profile: A CRLB-Based Optimality Analysis}
\author{
         Yixin Zhang, Yuzhou Li,~\IEEEmembership{Member,~IEEE}, Yu Zhang, and~Tao Jiang,~\IEEEmembership{Senior Member,~IEEE}

%\thanks{Parts of this work were presented at the ICT'13 and the IEEE ICC'14 and accepted by the IEEE GLOBECOM'14.}
\thanks{The authors are with the Wuhan National Laboratory for Optoelectronics and the School of Electronics Information and Communications, Huazhong University of Science and Technology, Wuhan, 430074, P. R. China (e-mail: \{zhangyixin, yuzhouli, yu\_zhang, taojiang\}@hust.edu.cn).}
}
\maketitle
\IEEEpeerreviewmaketitle
%---------------------------------------Make Abstract--------------------------------------
\begin{abstract}
Existing works have explored the anchor deployment for autonomous underwater vehicles (AUVs) localization under the assumption that the sound propagates straightly underwater at a constant speed. Considering that the underwater acoustic waves propagate along bent curves at varying speeds in practice, it becomes much more challenging to determine a proper anchor deployment configuration. In this paper, taking the practical variability of underwater sound speed into account, we investigate the anchor-AUV geometry problem in a 3-D time-of-flight (ToF) based underwater scenario from the perspective of localization accuracy. To address this problem, we first rigorously derive the Jacobian matrix of measurement errors to quantify the Cramer-Rao lower bound (CRLB) with a widely-adopted isogradient sound speed profile (SSP).
We then formulate an optimization problem that minimizes the trace of the CRLB subject to the angle and range constraints to figure out the anchor-AUV geometry, which is multivariate and nonlinear and thus generally hard to handle.
For mathematical tractability, by adopting tools from the estimation theory, we interestingly find that this problem can be equivalently transformed into a more explicit univariate optimization problem. By this, we obtain an easy-to-implement anchor-AUV geometry that yields satisfactory localization performance, referred to as the uniform sea-surface circumference (USC) deployment. Extensive simulation results validate our theoretical analysis and show that our proposed USC scheme outperforms both the cube and the random deployment schemes in terms of localization accuracy under the same parameter settings.
\end{abstract}
%--------------------------------------------- Make Key words-----------------------------
\section{Introduction}
\begin{figure*}[t]
\centering \leavevmode \epsfxsize=5.0 in  \epsfbox{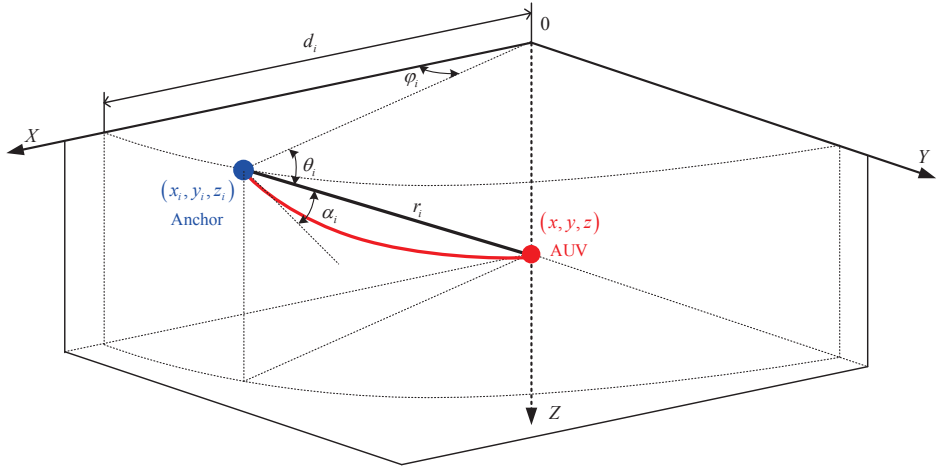}
\centering \caption{Illustration of the acoustic ray between the AUV and an anchor.} \label{Fig:Wave_propagation_path}
\end{figure*}

As an important platform for underwater information gathering and transmitting, autonomous underwater vehicles (AUVs) have been widely used for a variety of underwater applications such as oceanographic surveys, object detection, and environment monitoring
\cite{AUVNavigationandLocalizationReviewJOE2014,ResearchAndDevelopmentOfSoftwareSystemForLBLPrecisePositioningTheory/WU2013,AbsolutePositioningOfAnAutonomousUnderwaterVehicleUsingGPSAndAcousticMeasurements/JOE2005}.
In these applications, the observed data becomes meaningful only when labeled with correct position \cite{MarineBigData/CM2017,ModelingAUVLocalizationErrorInALongBaselineAcousticPositioningSystem/JOE2017,RobustLocalizationUsingRangeMeasurementsWithUnknownAndBoundedErrors/TWC2017,UltraDenseHetNetsMeetBigDateGreenFrameworksTechniquesAndApproaches/CM2017}, and thus how to accurately locate AUVs becomes a crucial issue.
However, universal underwater localization is not easy and still vacant currently, as the global positioning system (GPS) signals are unavailable in the ocean \cite{ModelingAUVLocalizationErrorInALongBaselineAcousticPositioningSystem/JOE2017,AbsolutePositioningOfAnAutonomousUnderwaterVehicleUsingGPSAndAcousticMeasurements/JOE2005}.
The most prominent choice for AUV localization is to use acoustic anchors deployed beforehand with known positions, such as surface buoys, seabed array elements, and sensors attached to autonomous surface vehicles (ASVs).
In these scenarios, the geometry relationships between the AUV and anchors, including both the anchor-AUV geometrical shape and ranges, have significant impacts on the localization network's overall performance, e.g., localization accuracy, network connectivity, seamless coverage, and energy efficiency \cite{ArrayElementLocalizationForHorizontalArraysViaOccamsInversion/JASA1998,OptimalityAnalysisofSensorTargetLocalizationGeometries/Automatics2010,ImpactsOfDeploymentStrategiesOnLocalizationPerformanceInUnderwaterAcousticSensorNetworks/TIE2015,PreciseThreeDimensionalSeafloorGeodeticDeformationMeasurementsUsingDifferenceTechniques/EPS2005,EnergyEfficientSubcarrierAssignmentAndPowerAllocationInOFDMASystemsWithMaxMinFairnessGuarantees/TCOM2015,OptimalArrayElementLocalization/JASA1999,TrackDesignForTheAcousticSensorInstallationAlignmentCalibrationInUltraShortBaselinePositioningSystem/SEE2016}.
In particular, this article focuses on designing proper anchor-AUV geometries for AUV localization from the perspective of localization accuracy.

There have been some works to investigate anchor-target geometries for target localization in underwater environments so far.
Han \emph{et al}. in
\cite{ImpactsOfDeploymentStrategiesOnLocalizationPerformanceInUnderwaterAcousticSensorNetworks/TIE2015}
proposed a regular tetrahedron geometry for anchor deployment, which outperforms the  random and cube deployment schemes in terms of network connectivity but achieves little improvement in localization accuracy.
In \cite{OptimalArrayElementLocalization/JASA1999}, Dosso \textit{et al.} studied the problem of determining the optimal configurations for symmetrically horizontal or vertical source placement in array element localization scenarios.
For high accuracy, N. Bishop \emph{et al}. in \cite{OptimalAnalysisOfTOA/ICIS2007} exploited passive time-of-arrival (ToA) based localization to explore the relative sensor-target geometries and proved that the way of equi-angularly surrounding the target by an arbitrary number of sensors is optimal.
In the case when the sensor-target distance is fixed, the sensor deployment for underwater source localization was analyzed based on the Cramer-Rao lower bound (CRLB) theory in \cite{SensorPlacementForUnderwaterSourceLocalizationWithFixedDistances/GRSL2016}.
Further in \cite{OptimalSensorPlacement/Underwater/Range-only/JOE2016}, by minimizing the trace of the CRLB, i.e., $\text{tr}(\text{CRLB})$, an optimal anchor deployment for underwater localization was derived under the assumption that the sound speed is constant.

In a nutshell, although \cite{ImpactsOfDeploymentStrategiesOnLocalizationPerformanceInUnderwaterAcousticSensorNetworks/TIE2015,OptimalAnalysisOfTOA/ICIS2007,OptimalArrayElementLocalization/JASA1999,SensorPlacementForUnderwaterSourceLocalizationWithFixedDistances/GRSL2016,OptimalSensorPlacement/Underwater/Range-only/JOE2016} investigated the anchor geometry deployment problem for underwater target localization from different perspectives, they all assumed that the speed of acoustic waves remains unchanged in underwater environments and their trajectories are straight lines. However, the underwater acoustic waves propagate along bent curves at varying speeds in practice due to the fact that the water medium is inhomogeneous in temperature, pressure, and salinity.
This unique phenomenon disrupts the linear dependency between the sound propagation distance and the duration (i.e., consumed time) it takes for the sound to travel over that distance.
Therefore, the existing straight-line propagation model based anchor deployment schemes
\cite{ImpactsOfDeploymentStrategiesOnLocalizationPerformanceInUnderwaterAcousticSensorNetworks/TIE2015,OptimalAnalysisOfTOA/ICIS2007,OptimalArrayElementLocalization/JASA1999,SensorPlacementForUnderwaterSourceLocalizationWithFixedDistances/GRSL2016,OptimalSensorPlacement/Underwater/Range-only/JOE2016}
are possibly not effective in realistic underwater environments as what we expect.
To address this problem, some works have begun to develop localization algorithms to match the feature of the varying sound speed, in which the acoustic velocity is modeled as the sound speed profile (SSP) in
\cite{ModelingAUVLocalizationErrorInALongBaselineAcousticPositioningSystem/JOE2017,ErrorEvaluationInAcousticPositioningOfASingleTransponderForSeafloorCrustalDeformationMeasurements/EPS2002,TargetLocalization/ToF/SSP/TSP2013,AbsolutePositioningOfAnAutonomousUnderwaterVehicleUsingGPSAndAcousticMeasurements/JOE2005,MotionCompensatedAcousticLocalizationorUnderwaterVehicles/JOE2016,AccurateShallowandDeepWaterRangEstimationforUnderwaterNetworks/GLOBALCOM2015,AnalyticalStudyOfAcousticRangingAccuracy/IEEE/OES2016,StratificationEffectCompensationforImprovedUnderwaterAcousticRanging/TSP2008,LocalizationUsingRayTracingforUnderwaterAcousticSensorNetworks/CL2010} or a series of state variables in \cite{KalmanBasedUnderwaterTrackingWithUnknownEffectiveSoundVelocity/IPOC2016,EffectOnKalmanBasedUnderwaterTrackingDueToOceanCurrentUncertainty/PAUV2016}. Both simulation results and field tests have demonstrated that localization algorithms that consider the varying sound speed can achieve higher localization accuracy than those based on a constant sound speed. Nevertheless, \cite{ModelingAUVLocalizationErrorInALongBaselineAcousticPositioningSystem/JOE2017,ErrorEvaluationInAcousticPositioningOfASingleTransponderForSeafloorCrustalDeformationMeasurements/EPS2002,TargetLocalization/ToF/SSP/TSP2013,AbsolutePositioningOfAnAutonomousUnderwaterVehicleUsingGPSAndAcousticMeasurements/JOE2005,MotionCompensatedAcousticLocalizationorUnderwaterVehicles/JOE2016,AccurateShallowandDeepWaterRangEstimationforUnderwaterNetworks/GLOBALCOM2015,AnalyticalStudyOfAcousticRangingAccuracy/IEEE/OES2016,StratificationEffectCompensationforImprovedUnderwaterAcousticRanging/TSP2008,LocalizationUsingRayTracingforUnderwaterAcousticSensorNetworks/CL2010,KalmanBasedUnderwaterTrackingWithUnknownEffectiveSoundVelocity/IPOC2016,EffectOnKalmanBasedUnderwaterTrackingDueToOceanCurrentUncertainty/PAUV2016}
obtained their results under some specialized anchor deployment schemes without providing any analysis or proof for the problem whether such deployments are appropriate or not.

In view of these, taking the practical variability of underwater sound speed into account, this paper devotes to answering whether there exist more proper deployment configurations that can further improve the localization accuracy. We explore this fundamental problem by considering a 3-D time-of-flight (ToF) \cite{LocalizationAlgorithmsAndStrategiesForWirelessSensorNetworks} based underwater localization scenario involving multiple anchors and a single target AUV.
Under a realistic isogradient SSP, we first derive the CRLB of ToF-based estimation methods as the localization accuracy index to evaluate whether a spatial anchor configuration is appropriate or not. By minimizing the trace of CLRB, we formulate an optimization problem to figure out the promising anchor-AUV geometry. To handle this complicated problem, we first convert it into another tractable form subject to some angle and range constraints.
We then find an easy-to-implement anchor-AUV geometrical shape that satisfies these constraints and obtain corresponding anchor-AUV ranges. Simulation results show that our proposed anchor-AUV geometry achieves higher localization accuracy when targeting the AUV.

The main contributions of this work are as follows.

\begin{itemize}
    \item We first rigorously derive the Jacobian matrix of measurement errors evaluated at the AUV's true position to quantify the CRLB of the ToF-based AUV localization under an isogradient SSP for characterizing the localization accuracy with a given anchor-AUV geometry.
        Enfolding the variable sound speed into the anchor geometry deployment problem, we then formulate an anchor-AUV geometry optimization problem subject to the angle and range constraints by minimizing the trace of the CRLB.
        As this problem is rather complex to deal with mathematically, we adopt the Courant-Fischer-Weyl min-max principle and the arithmetic mean and geometric mean (AM-GM) inequality to transform it equivalently into a more explicit optimization problem.
    \item Although it is in general difficult to find a globally optimal solution for our formulated nonlinear and multivariate geometry problem, we obtain an easy-to-implement anchor-AUV geometry that guarantees satisfactory localization accuracy in a semi-analytical way, referred to as the uniform sea-surface circumference (USC) deployment. Specifically, we first devise an anchor-AUV geometrical shape to convert the original multivariate problem into a univariate distance optimization problem. We then adopt a gradient descent method to find the promising anchor-AUV ranges given the USC deployment scheme.
    \item Extensive simulation results validate our theoretical analysis and show that our proposed USC scheme has a better localization performance compared with the existing cube and the random deployment schemes under the same parameter settings. Moreover, the USC deployment can maintain a satisfactory localization accuracy even if the relative anchor-AUV geometry changes, which indicates that the USC is robust to position variance of the AUV and anchors.
\end{itemize}

The remainder of this paper is organized as follows. In Section \uppercase\expandafter{\romannumeral2}, we introduce the system model and formulate an anchor deployment optimization problem. The USC deployment is devised in Section \uppercase\expandafter{\romannumeral3}. Section \uppercase\expandafter{\romannumeral4} presents simulation results to evaluate the performance of the USC. Finally, we conclude our paper in Section \uppercase\expandafter{\romannumeral5}.
\section{System Model and Problem Formulation}
In this section, we first introduce the considered system scenario and involved parameters. Then, we adopt the isogradient sound speed profile to quantify the localization error in terms of the CRLB and use it to formulate the anchor-AUV geometry optimization problem.
\subsection{System Scenario} \label{Subsection:SystemScenario}
%\mathbb{R}
As shown in Fig.~\ref{Fig:Wave_propagation_path}, we consider a 3-D underwater acoustic localization scenario, where a target AUV is to be located with the assistance of $N$ similar anchors.
In this ${\Re}^3$ scenario, we denote the set of anchors by $\mathcal{C}=\{1,2,3,...,N\}$, the location of the AUV by $ {\textbf{q}} = {\left[ {x,y,z} \right]^{\rm{T}}}$, and the position of the $i$th anchor by ${\textbf{p}_i} = {\left[ {x_i,y_i,z_i} \right]^{\rm{T}}}$, $ \forall i \in \mathcal{C}$. Let the range (i.e., distance) between the $i$th anchor and the AUV be $r_i=\| {\textbf{q}}-{\textbf{p}}_i\|$, where $\|\cdot\|$ denotes the Euclidean norm.

Similarly as in \cite{AcousticModels/JOE1993}, assume that the underwater sound speed follows an isogradient depth-dependent SSP and has the following form
\begin{equation}\label{Eq:SoundSpeedProfile}
 C\left( z \right) = b + az
\end{equation}
where $b$ denotes the sound speed at the sea surface, and $a$ is the steepness of the SSP.
Because of the variable sound speed, the actual acoustic path from anchor $i$ to the AUV follows a bent trajectory, as depicted by the red bold curve in Fig.~\ref{Fig:Wave_propagation_path}.
The angle at which the actual ray trajectory deviates from the straight line between the AUV and anchor $i$, denoted by $\alpha_i$, can be derived from a set of differential equations characterized by the Snell's law and can be calculated as \cite{TargetLocalization/ToF/SSP/TSP2013}
\begin{equation}\label{Df:Alpha_i}
{\alpha _i} = \arctan \frac{{\sqrt {{{\left( {x - {x_i}} \right)}^2} + {{\left( {y - {y_i}} \right)}^2}} }}{{\frac{{2b}}{a} + \left( {z + {z_i}} \right)}}
\end{equation}
and the angle of this straight line w.r.t the horizontal axis, denoted by $\theta_i$, is equal to
\begin{equation}\label{Df:Theta_i}
{\theta _i} = \arctan \frac{{z - {z_i}}}{{\sqrt {{{\left( {x - {x_i}} \right)}^2} + {{\left( {y - {y_i}} \right)}^2}} }}.
\end{equation}

In addition, the horizontal distance and the horizontal orientation angle between the $i$th anchor and the AUV are calculated as ${d_i} = \sqrt {{{\left( {x - {x_i}} \right)}^2} + {{\left( {y - {y_i}} \right)}^2}}$ and ${\varphi _i} = \arctan \frac{{y - {y_i}}}{{x - {x_i}}}$, respectively.
With these information about the anchor-AUV geometry, the actual ray traveling time (RTT) from anchor $i$ to the AUV along the bent curve is derived as  \cite{TargetLocalization/ToF/SSP/TSP2013}
\begin{equation}\label{Eq:RayTravelingTime}
\begin{aligned}
t_i = \frac{1}{a}\left[ {\ln \left( {\frac{{1 + \sin \left( {{\theta _i} + {\alpha _i}} \right)}}{{\cos \left( {{\theta _i} + {\alpha _i}} \right)}}} \right) - \ln \left( {\frac{{1 + \sin \left( {{\theta _i} - {\alpha _i}} \right)}}{{\cos \left( {{\theta _i} - {\alpha _i}} \right)}}} \right)} \right].
\end{aligned}
\end{equation}
From Eqs.~(\ref{Df:Alpha_i})--(\ref{Eq:RayTravelingTime}), it is noteworthy that the RTT is dependent not only on the distance between the AUV and anchors, but also on their positions. This feature makes the underwater localization different from the terrestrial localization as the RTT in the terrestrial is only related to the anchor-target range.

In a ToF-based localization system, the position of the AUV is estimated by using at least four RTT measurements from four anchors.
The RTT measurement from the anchor $i$ to the AUV, denoted by ${\hat {t}}_i$, can be modeled as
\begin{equation}\label{Eq:HatT_i}
{{\hat {t}}_i}={t_i} + {\xi_i},~ \forall i \in \mathcal{C}
\end{equation}
where ${\xi_i}$ refers to the measurement error, which is assumed to be mutually independent and Gaussian distributed with zero mean and distance-dependent variance $\sigma _i^2$ \cite{TDOABasedSourceLocalizationWithDistanceDependentNoises/TWC2015}, i.e., $\xi_i\sim{\cal N}(0,\sigma _i^2)$.
For notational simplicity, we use $\textbf n =[{\xi_1},...,{\xi_N}]^{\rm T}$ to denote the measurement noise and $\textbf{v}(\textbf{q})=[{t_1},...,{t_N}]^{\rm T}$ to denote the actual RTT with respect to the AUV's position $\textbf q$.
Thus, the RTT measurements between the AUV and $N$ anchors, denoted by $\hat{\textbf{v}}$, can be written in a vector format as
\begin{equation}\label{Eq:MeasurementVector}
\begin{aligned}
\hat {\textbf{v}}= \textbf{v}( {\textbf{q}})+\textbf{n}={\left[ {{t_1},...,{t_N}} \right]^{\rm{T}}} + {\left[ {{\xi_1},...,{\xi_N}} \right]^{\rm{T}}}.
\end{aligned}
\end{equation}
The measurement error $\textbf{n}$ is assumed to be a multivariate random vector with an $N\times N$ positive-definite covariance matrix \cite{StatisticalTheroy/Location/TAE1984}, denoted as
\begin{equation}\label{Eq:MeasurementConvarienceMatrix}
{\bf{\Sigma }}  = {\mathbb{E}}\{(\textbf{n}-{\textbf{E}(\textbf{n})}){{(\textbf{n}-{\textbf{E}(\textbf{n})})}^{\rm{T}}}\}
\end{equation}
where ${\mathbb{E}}\{\cdot\}$ denotes the expectation operation. By substituting (\ref{Eq:MeasurementVector}) into (\ref{Eq:MeasurementConvarienceMatrix}), the covariance matrix $\bf{\Sigma }$ can be derived as
\begin{equation}
{\bf{\Sigma}} =
\left[ {\begin{array}{*{20}{c}}
{\sigma _1^2}&{}&{}\\
{}& \ddots &{}\\
{}&{}&{\sigma _N^2}
\end{array}} \right].
\end{equation}

\par\noindent The list of symbols and notations used throughout this paper is summarized in Table
\ref{Table:SymbolsandDescription}.
\begin{table}[t]
\centering
\caption{\label{Table:SymbolsandDescription}Symbols and Descriptions}
\begin{supertabular}{l >{}p{5.3cm}}
\shline
Symbol & Description\\
\hline
$\mathcal{C}$ & Set of anchors\\
\hline
$i$ & Index of the anchor\\
\hline
$N$ & Number of anchors\\
\hline
$\textbf q = {\left[ {x,y,z} \right]^{\text T}}$ & Location of the AUV\\
\hline
${{\textbf p}_i} = {\left[ {{x_i},{y_i},{z_i}} \right]^{\text T}}$ & Location of the $i$th anchor\\
\hline
${\theta_i}$ & Angle of the straight line between the $i$th anchor and the AUV w.r.t the horizontal axis\\
\hline
${\alpha_i}$ & Angle at which the ray trajectory deviates from the straight line between the $i$th anchor and the AUV\\
\hline
${\varphi _i}$ & Horizontal orientation angle between the $i$th anchor and the AUV\\
\hline
$d_i$ & Horizontal distance between the $i$th anchor and the AUV\\
\hline
$r_i$ & Range between the $i$th anchor and the AUV\\
\hline
$t_i$ & Actual ray traveling time (RTT) from the $i$th anchor to the AUV\\
\hline
${\sigma _i^2}$ & Measurement noise variance of the RTT $t_i$\\
\hline
${\bf{\emph{J}}_o}$ & Jacobian of measurement errors evaluated at the AUV's true position\\
\hline
${\bf{\Sigma }} $ & Measurement covariance matrix of anchors\\
\hline
\shline
\end{supertabular}
\end{table}

\newcounter{mytempeqncnt2}
\setcounter{mytempeqncnt2}{\value{equation}}
\setcounter{equation}{12}
\begin{figure*}[hb]
\hrulefill % 一条线
\begin{equation}\label{Eq:FIMofToFandSSP}
{\text {FIM}} = \frac{4}{{{a^2}}}\left[ {\begin{array}{*{20}{c}}
{\sum\limits_{i = 1}^N {\frac{{{{\sin }^2}{\alpha _i}{{\cos }^2}{\theta _i}{{\cos }^2}{\varphi _i}}}{{d_i^2\sigma _i^2}}} }&{\sum\limits_{i = 1}^N {\frac{{{{\sin}^2} {\alpha _i}{{\cos }^2}{\theta _i}\sin{\varphi _i}\cos {\varphi _i}}}{{d_i^2\sigma _i^2}}} }&{\sum\limits_{i = 1}^N {\frac{{{{\sin }^2}{\alpha _i}\tan \left( {{\theta _i} - {\alpha _i}} \right){{\cos }^2}{\theta _i}\cos {\varphi _i}}}{{d_i^2\sigma _i^2}}} }\\
{\sum\limits_{i = 1}^N {\frac{{{{\sin}^2} {\alpha _i}{{\cos }^2}{\theta _i}\sin{\varphi _i}\cos {\varphi _i}}}{{d_i^2\sigma _i^2}}} }&{\sum\limits_{i = 1}^N {\frac{{{{\sin }^2}{\alpha _i}{{\cos }^2}{\theta _i}{{\sin }^2}{\varphi _i}}}{{d_i^2\sigma _i^2}}} }&{\sum\limits_{i = 1}^N {\frac{{{{\sin}^2} {\alpha _i}\tan \left( {{\theta _i} - {\alpha _i}} \right){{\cos }^2}{\theta _i}\sin {\varphi _i}}}{{d_i^2\sigma _i^2}}} }\\
{\sum\limits_{i = 1}^N {\frac{{{{\sin }^2}{\alpha _i}\tan \left( {{\theta _i} - {\alpha _i}} \right){{\cos }^2}{\theta _i}\cos {\varphi _i}}}{{d_i^2\sigma _i^2}}} }&{\sum\limits_{i = 1}^N {\frac{{{{\sin}^2} {\alpha _i}\tan \left( {{\theta _i} - {\alpha _i}} \right){{\cos }^2}{\theta _i}\sin {\varphi _i}}}{{d_i^2\sigma _i^2}}} }&{\sum\limits_{i = 1}^N {\frac{{{{\sin }^2}{\alpha _i}{{\tan }^2}\left( {{\theta _i} - {\alpha _i}} \right){{\cos }^2}{\theta _i}}}{{d_i^2\sigma _i^2}} } }
\end{array}} \right]
\end{equation}
\end{figure*}
\setcounter{equation}{\value{mytempeqncnt2}}

In this paper, we devote to constructing a geometrical deployment configuration between the AUV and anchors for improving localization accuracy, i.e., determining the aforementioned ranges and angles (namely $d_i$, $r_i$, $\alpha_i$ in (\ref{Df:Alpha_i}), $\theta_i$ in (\ref{Df:Theta_i}), and $\varphi_i$).
Different configurations achieve different localization performance, and thus exploring the proper geometry is equivalent to finding the angular and distance relations. In next subsection, we will first derive the localization performance bound by using all above angular and distance formulas. Based on the derived bound, we then formulate an optimization problem to optimize angular and distance configuration, which forms the proper anchor-AUV geometry.

\subsection{Problem Formulation} \label{Subsection:TheCramer-RaoLowerBoundFisherInformationMatrix}
The performance of an estimator is commonly characterized by the mean-square error (MSE) of the estimate, which represents the uncertainty associated with the accuracy of the estimation results. One of classical results for this is known as the Cramer-Rao lower bound (CRLB), which has been seen to provide a reasonably tight bound on the MSE of the estimate \cite{OptimalMeasurementMethodsForDistributedParameterSystemIdentification/CRCPress2005}. Thus, the performance of localization, which is essentially an estimation problem, can be explicitly depicted by the CRLB of the estimator. Further, motivated by the fact that the anchor deployment geometry has significant impacts on the performance of localization estimators, we in this subsection derive the performance bound, i.e., the CRLB, to evaluate whether a relative anchor-AUV geometry is appropriate or not.

Back to our considered 3-D ToF-based AUV localization scenario (see Fig.~\ref{Fig:Wave_propagation_path}), the AUV's true position $\textbf q  = {\left[ {x,y,z} \right]^{\rm{T}}} \in {{\Re}^3}$ needs to be estimated, given the observation measurement $\hat{\textbf v}$. It is known that, the Cramer-Rao inequality bounds the achievable covariance by an unbiased estimator. For an unbiased estimate ${\hat {\textbf{q}}}$ of ${{\textbf{q}}}$, the CRLB states that \cite{AccuratePassiveLocationEstimationUsingTOAMeasurements/TWC2012,OptimalMeasurementMethodsForDistributedParameterSystemIdentification/CRCPress2005}
\setcounter{equation}{8}
\begin{equation}\label{Df:FIMandCRLB}
{\mathbb{E}}\{ {\left( {\hat {\textbf{q}} - {\textbf{q}}} \right){{\left( {\hat {\textbf{q}} - {\textbf{q}}} \right)}^{\rm{T}}}} \} \ge {\text{FIM}^{ - 1}} \buildrel \Delta \over = {\text{CRLB}}
\end{equation}
where the FIM is called the Fisher Information Matrix (FIM), an index quantifying the information amount that the observable random measurement vector $\hat {\textbf{v}}$ carries about the unobservable $\textbf{q}$.

Under the standard assumption of Gaussian measurement errors, the probability function of $\textbf{q}$, given the measurement vector $ \hat{\textbf{v}}\sim{\cal N}(\textbf{v}(\textbf{q}),{\bf{\Sigma }})$, is given by
\begin{equation}\label{Df:LikelihoodFunction}
%\begin{small}
{{\textbf{f}}_{\textbf{q}} }= \frac{1}{{{{\left( {2\pi } \right)}^{\frac{N}{2}}}{{\left| {\bf{\Sigma }}  \right|}^{\frac{1}{2}}}}}\exp \left\{ { - \frac{1}{2}{{\left( {\hat {\textbf{v}} - \textbf{v}\left( {\textbf{q}} \right)} \right)}^{\rm{T}}}{{\bf{\Sigma }}^{ - 1}}\left( {\hat { \textbf{v}} - \textbf{v}\left( {\textbf{q}} \right)} \right)} \right\}
%\end{small}
\end{equation}
where $|{\bf{\Sigma }}|$ is the determinant of ${\bf{\Sigma }}$.
As in  \cite{StatisticalTheroy/Location/TAE1984} and \cite{RSSBasedLocationEstimationWithUnknownPathlossModel/TWC2006}, by taking the logarithm on (\ref{Df:LikelihoodFunction}), computing its derivative with respect to ${\textbf{q}}$, and taking the expectation operation, the FIM corresponding to our considered scenario can be derived as
\begin{equation}\label{Df:FisherInformationMatrixDefination}
%\begin{small}
{\text{FIM}}  = {\mathbb{E}}\left\{ {{\nabla _{\textbf{q}}}\log {{\textbf{\textit{f}}}_{\textbf{q}}} \cdot {\nabla _{\textbf{q}}}\log {{\textbf{\textit{f}}}_{\textbf{q}}}^{\rm{T}}} \right\} = {\bf{\emph{J}}}_o^{\rm{T}}{{\bf{\Sigma }} ^{ - 1}}{{\bf{\emph{J}}}_o}.
%\end{small}
\end{equation}

\par\noindent In (\ref{Df:FisherInformationMatrixDefination}), ${\bf{\emph{J}}}_o \in {\Re ^{{N}  \times 3}}$ is the Jacobian of measurement vector with respect to
${\textbf{q}}$, which is specified in the following theorem, proved in Appendix~\ref{Appendix:ProofOfJacobianMatrix}.
\begin{theorem} \label{Theorem:JacobianMatrix}
Considering the 3-D ToF-based localization problem under an isogradient SSP, the Jacobian of measurement errors evaluated at the AUV's true position $\textbf q$ is given by
\begin{equation}\label{Df:JacobianMatrix}
{\bf{\emph{J}}}_o = {\frac{2}{a}}\left[ {\begin{array}{*{20}{c}}
{\frac{{\sin {\alpha _1}\cos {\varphi _1}}}{{{r_1}}}}&{\frac{{\sin {\alpha _1}\sin {\varphi _1}}}{{{r_1}}}}&{\frac{{\tan \left( {{\theta _1} - {\alpha _1}} \right)\sin {\alpha _1}}}{{{r_1}}}}\\
 \vdots & \vdots & \vdots \\
{\frac{{\sin {\alpha _i}\cos {\varphi _i}}}{{{r_i}}}}&{\frac{{\sin {\alpha _i}\sin {\varphi _i}}}{{{r_i}}}}&{\frac{{\tan \left( {{\theta _i} - {\alpha _i}} \right)\sin {\alpha _i}}}{{{r_i}}}}\\
 \vdots & \vdots & \vdots \\
{\frac{{\sin {\alpha _N}\cos {\varphi _N}}}{{{r_N}}}}&{\frac{{\sin {\alpha _N}\sin {\varphi _N}}}{{{r_N}}}}&{\frac{{\tan \left( {{\theta _N} - {\alpha _N}} \right)\sin {\alpha _N}}}{{{r_N}}}}
\end{array}} \right].
\end{equation}
\end{theorem}

From Theorem~\ref{Theorem:JacobianMatrix}, by substituting (\ref{Eq:MeasurementConvarienceMatrix}) and (\ref{Df:JacobianMatrix}) into (\ref{Df:FisherInformationMatrixDefination}), we can obtain the full expression of the FIM as shown in (\ref{Eq:FIMofToFandSSP}) at the bottom of this page, and thus the CRLB could be derived by ${\text{CRLB = FI}}{{\text{M}}^{ - 1}}$ from (\ref{Df:FIMandCRLB}).
There are several optimality criteria that can be considered, e.g., the A-optimality, the D-optimality, and the E-optimality criteria \cite{StatisticalTheroy/Location/TAE1984}. The D-optimality criterion aims to maximize the determinant of the FIM or to minimize the volume of the localization error ellipsoid. However, it can yield to some errors, as the information in one dimension can be improved rapidly, providing a very large FIM determinant, while we may not have any information in other dimensions \cite{OptimalSensor/AOA/Aoptimal/sensors2013}. This problem can be avoided with the A- and E- optimality criteria which aim to minimize the trace and the maximum eigenvalue of the CRLB, respectively.
Among them, the A-optimality is widely adopted in existing works, e.g., \cite{OptimalSensor/AOA/Aoptimal/sensors2013,OptimalSensorPlacementFor3DAoATargetLocalization/2017/TAE,OptimalSenorDeploymentFor3DAOATargetLocalization/ICASSP2015}, and it is more appropriate in our problem.
%Furthermore,  \cite{OptimalSensor/AOA/Aoptimal/sensors2013,StatisticalTheroy/Location/TAE1984}  have proved that the trace of CRLB, i.e., tr(CRLB), is a reasonable index to evaluate the accuracy of target position estimation and showed that a lower value of the tr(CRLB) indicates a better estimation result and vice versa.
Based on this fact, we adopt the tr(CRLB) as the performance indicator for the AUV localization in this paper. Specifically, for a fixed AUV location, we formulate the following tr(CRLB) minimization problem\footnote{It is noteworthy that changing the minimization objective in problem (\ref{Op:MinimizationProblem}) as a min-max objective is more interesting when considering that the AUV is moving inside a bounded operation area. Although our analysis is based on the fixed AUV scenario, we will verify that our proposed scheme can also achieve satisfactory localization performance in a moving AUV scenario through the simulation result in Section \uppercase\expandafter{\romannumeral4}.}
\setcounter{equation}{13}
\begin{equation}\label{Op:MinimizationProblem}
\begin{aligned}
{\mathop {\min }_{{\alpha_i}, {\theta_i}, {\varphi_i}, {d_i}}}~~& \text{tr(CRLB)}\\
\text{s.t.}   ~~&\text{C1:} ~~{\alpha _i} \in \left[ {0,\frac{\pi }{2}} \right),~\forall i \in \mathcal{C}\\
              ~~&\text{C2:} ~~{\theta _i} \in \left[ {0,\frac{\pi }{2}}\right),~\forall i \in \mathcal{C} \\
              ~~&\text{C3:} ~~{\varphi_i} \in \left[ {0,2\pi }\right],~\forall i \in \mathcal{C} \\
              ~~&\text{C4:} ~~d_i>0,~\forall i \in \mathcal{C}\\
              %~~&\text{C5:} ~~r_i>0,~\forall i \in \mathcal{C}\\
              ~~&\text{C5:} ~~N\geq4.
\end{aligned}
\end{equation}
\par\noindent Regarding problem (\ref{Op:MinimizationProblem}), angles $\alpha _i$, $\theta_i$, and $\varphi_i$ determine the geometrical shape between the anchors and the AUV, while $d_i$ denotes the anchor-AUV horizontal distances. C1--C4 directly follow the geometric logic, and C5 comes from the fact that at least four anchors are needed to confirm the AUV's position in 3-D underwater scenarios.

\section{Anchor Deployment Solutions}
\label{Section:OptimalAnchorDeploymentSolutions}
To solve (\ref{Op:MinimizationProblem}), a straightforward method is to first derive the closed-form expression for the tr(CRLB) and then find its minimum, which is in general difficult due to the fact that
the CRLB is the inverse of the complicated matrix (\ref{Eq:FIMofToFandSSP}).
As a result, there are usually no easy-to-calculate solutions for the
CRLB as functions of the anchor-AUV angles and distances.
Aiming at solving (\ref{Op:MinimizationProblem}) in a more explicit manner, we exploit a classical result in the Courant-Fischer-Weyl min-max principle \cite{OptimalSensor/AOA/Aoptimal/sensors2013} to reformulate it, described in the following theorem. We refer the readers to \cite{OptimalSensor/AOA/Aoptimal/sensors2013} on its proof and omit it for brevity in the paper.
\begin{theorem} \label{Theorem:CourantFischerWeyl}
The {\rm{FIM}} in a 3-D target localization scenario is a $3\times3$ symmetric matrix and can be expressed as
\begin{equation}
\rm{FIM}  = \left[ {\begin{array}{*{20}{c}}
{{\psi _{11}}}&{{\psi _{12}}}&{{\psi _{13}}}\\
{{\psi _{21}}}&{{\psi _{22}}}&{{\psi _{23}}}\\
{{\psi _{31}}}&{{\psi _{32}}}&{{\psi _{33}}}
\end{array}} \right]
\end{equation}
where ${\psi _{12}}={\psi _{21}}$, ${\psi _{13}}={\psi _{31}}$, and ${\psi _{23}}={\psi _{32}}$. Moreover, the trace of the {\rm{CRLB}} is always larger than the sum of the {\rm{FIM}}'s diagonal elements, that is,
\begin{equation}
{\rm{tr}}\left( {{\rm{CRLB}}} \right) = {\rm{tr}}\left( {{\rm{FIM} ^{ - 1}}} \right) \ge \frac{1}{{{\psi _{11}}}} + \frac{1}{{{\psi _{22}}}} + \frac{1}{{{\psi _{33}}}}
\end{equation}
with the equality holding if and only if
\begin{equation}
{\psi _{12}} = {\psi _{13}} = {\psi _{23}}=0.
\end{equation}
\end{theorem}
\begin{remark}\label{Rm:Remark1}
Theorem~\ref{Theorem:CourantFischerWeyl} indicates that minimizing the $\rm{tr(CRLB)}$ is equivalent to minimizing $\frac{1}{{{\psi _{11}}}} + \frac{1}{{{\psi _{22}}}} + \frac{1}{{{\psi _{33}}}}$ subject to the constraint that the \rm{FIM} is diagonal, i.e., ${\psi _{12}} = {\psi _{13}} = {\psi _{23}}=0$.
\end{remark}

By applying Theorem~\ref{Theorem:CourantFischerWeyl} to the FIM matrix (\ref{Eq:FIMofToFandSSP}), problem (\ref{Op:MinimizationProblem}) can be equivalently transformed to
\begin{equation}\label{Op:SecondaryMinTrCRLB}
\begin{small}
\begin{aligned}
{\mathop {\min }_{{\alpha_i}, {\theta_i}, {\varphi_i}, {d_i}}}& {\text {tr}\left( \text {CRLB} \right) } ~~~~~~~~~~~~~~~~~\\
&=\frac{{{a^2}}}{{\sum\limits_{i = 1}^N {\frac{{4{{\sin }^2}{\alpha _i}{{\cos }^2}{\theta _i}{{\cos }^2}{\varphi _i}}}{{d_i^2\sigma _i^2}}} }} + \frac{{{a^2}}}{{\sum\limits_{i = 1}^N {\frac{{4{{\sin }^2}{\alpha _i}{{\cos }^2}{\theta _i}{{\sin }^2}{\varphi _i}}}{{d_i^2\sigma _i^2}}} }} \\
& +\frac{{{a^2}}}{{\sum\limits_{i = 1}^N {\frac{{4{{\tan }^2}\left( {{\theta _i} - {\alpha _i}} \right){{\cos }^2}{\theta _i}{{\sin }^2}{\alpha _i}}}{{d_i^2\sigma _i^2}} } }} \\
\text{s.t.} \ \ \  ~&\text{C1--C5}\\
              ~~&\text{C6:}\sum\limits_{i = 1}^N {\frac{{{\sin}^2} {\alpha _i}{{\cos }^2}{\theta _i}\sin {\varphi _i}\cos {\varphi _i}}{{d_i^2}{\sigma _i^2}}}  = 0 \\
              ~~&\text{C7:}\sum\limits_{i = 1}^N {\frac{{{\sin }^2}{\alpha _i}\tan \left( {{\theta _i} - {\alpha_i}} \right){{\cos }^2}{\theta _i}\cos {\varphi _i}}{{d_i^2}{\sigma _i^2}}}  = 0 \\
              ~~&\text{C8:} \sum\limits_{i = 1}^N {\frac{{{\sin}^2} {\alpha _i}\tan \left( {{\theta _i} - {\alpha _i}} \right){{\cos }^2}{\theta _i}\sin {\varphi _i}}{{d_i^2}{\sigma _i^2}}}  = 0.
\end{aligned}
\end{small}
\end{equation}

In (\ref{Op:SecondaryMinTrCRLB}), constraints C7--C9 form into a necessary condition to minimize the tr(CRLB), under which the FIM turns into a diagonal matrix and hence the expression of the tr(CRLB) becomes explicit to understand and deal with mathematically. Following the AM-GM inequality as defined and proved in \cite{PerformanceBoundsOnAverageErrorRatesUsingTheAMGMInequalityAndTheirApplicationsInRelayNetworks/TWC2012}, we further specify this necessary condition and clarify the tr(CRLB) minimization problem through the following lemma.

\begin{lemma}\label{Lemma:AMGM}
For any two non-negative numbers $a$ and $b$, the arithmetic mean of $a$ and $b$ is greater than or equal to the geometric mean of them, i.e.,
\begin{equation}\label{Eq:AMGM_Inequation}
\begin{aligned}
&\frac{1}{a} + \frac{1}{b} \ge \frac{2}{{\sqrt a \sqrt b }}\\
&\frac{1}{{2\sqrt a \sqrt b }} \ge \frac{1}{{a + b}}.
\end{aligned}
\end{equation}
Eq.~(\ref{Eq:AMGM_Inequation}) implies that
\begin{equation}\label{Eq:AMGM}
\frac{1}{a} + \frac{1}{b} \ge \frac{4}{{a + b}}
\end{equation}
where the equality holds if and only if $a=b$.
\end{lemma}
From Lemma~\ref{Lemma:AMGM}, respectively replacing $a$ and $b$ in (\ref{Eq:AMGM}) by ${\sum\limits_{i = 1}^N {\frac{{{\sin }^2}{\alpha _i}{{\cos }^2}{\theta_i}{{\cos }^2}{\varphi _i}}{{d_i^2}{\sigma _i^2}}} }$ and ${\sum\limits_{i = 1}^N {\frac{{{\sin }^2}{\alpha _i}{{\cos }^2}{\theta_i}{{\sin }^2}{\varphi _i}}{{d_i^2}{\sigma _i^2}}} }$, (\ref{Op:SecondaryMinTrCRLB}) can be further equivalently recast to
\begin{equation}\label{Op:Thrid}
\begin{small}
\begin{aligned}
{\mathop {\min }_{{\alpha_i}, {\theta_i}, {\varphi_i}, {d_i}}}& \text {tr(CRLB)}\\
&=\frac{{{a^2}}}{{\sum\limits_{i = 1}^N {\frac{{{{\sin }^2}{\alpha _i}}{{\cos }^2}{\theta_i}}{{d_i^2\sigma _i^2}}} }} + \frac{{{a^2}}}{{\sum\limits_{i = 1}^N {\frac{{4{{\tan }^2}\left( {{\theta _i} - {\alpha _i}} \right){{\sin }^2}{\alpha _i}}{{\cos }^2}{\theta_i}}{{d_i^2\sigma _i^2}} } }}\\
\text{s.t.} \ \ \ ~~&\text{C1--C8}\\
              ~~&\text{C9:}{\sum\limits_{i = 1}^N {\frac{{{\sin }^2}{\alpha _i}{{\cos }^2}{\theta_i}{{\cos }^2}{\varphi _i}}{{d_i^2}{\sigma _i^2}}} }
              ={\sum\limits_{i = 1}^N {\frac{{{\sin }^2}{\alpha _i}{{\cos }^2}{\theta_i}{{\sin }^2}{\varphi _i}}{{d_i^2}{\sigma _i^2}}} }.
\end{aligned}
\end{small}
\end{equation}

\newcounter{mytempeqncnt3}
\setcounter{mytempeqncnt3}{\value{equation}}
\setcounter{equation}{25}
\begin{figure*}[hb]
\hrulefill % 一条线
\begin{small}
\begin{equation}\label{Eq:CRLBEquality}
{\text{tr(CRLB)}} = \frac{{{a^2}{\sigma ^2}\left( {1 + {k^2}} \right)\left[ {{k^2}{z^2} + {{\left( {\frac{{2b}}{a} + z} \right)}^2}} \right]}}{{N{k^2}}} + \frac{{{a^2}}}{4}{\left[ {\frac{N}{{\left( {1 + {k^2}} \right)\left[ {{{\left( {\frac{{2b}}{a} + z} \right)}^2} + {{\left( {kz} \right)}^2}} \right]}} \cdot {{\left[ {\frac{{{k^2}z - \left( {\frac{{2b}}{a} + z} \right)}}{{\frac{{2b}}{a} + 2z}}} \right]}^2} \cdot \frac{1}{{{\sigma ^2}}} } \right]^{ - 1}}
\end{equation}
\end{small}
\end{figure*}
\setcounter{equation}{\value{mytempeqncnt3}}

From Problem (\ref{Op:Thrid}), to achieve possibly high estimation accuracy, the first and foremost requirement is that the anchor-AUV geometry should meet the constraints C1--C9 shown in (\ref{Op:MinimizationProblem}), (\ref{Op:SecondaryMinTrCRLB}), and (\ref{Op:Thrid}), where C1--C9 are a set of multivariable nonlinear equations. Although it is rather complex to mathematically quantify the complete solutions for C1--C9, we interestingly find that an easy-to-implement anchor-AUV geometrical shape, as a special solution for them, can be obtained by exploiting a semi-analytical method. The following theorem characterizes this geometrical shape.
\begin{theorem} \label{Theorem:ASpecialSolution}
One of solutions that meets C1--C9 is given by
\begin{equation} \label{Eq:ASpecialSolution}
\left\{
\begin{aligned}
~~&  \left| {{\alpha _1}} \right|  =  \cdots  = \left| {{\alpha _i}} \right| =  \cdots  = \left| {{\alpha _N}} \right|=\alpha \\
~~&  \left| {{\theta _1}} \right|  =  \cdots  = \left| {{\theta _i}} \right| =  \cdots  = \left| {{\theta _N}} \right|=\theta \\
~~& {\varphi _i} = \frac{{2\pi i}}{N}\\
~~& {d_1}  = ...={d_i}=... = {d_N}=d \\
~~& {\sigma _1}^2 = ... = {\sigma _i}^2 = .... = {\sigma _N}^2={\sigma}^2 \\
%~~& {r_1}  = ...={r_i}=... = {r_N}=r \\
~~& z_i=0
\end{aligned}
\right .
%, \forall i \in \left\{ {1,...,N \ge 4} \right\}.
\end{equation}
where $N\geq4$ and $\forall i \in \mathcal{C}$.
\end{theorem}
\begin{IEEEproof}\label{Proof:ASpecialSolution}
When $N \geq 4$, noticing the orthogonality relations for sines and cosines from Fourier analysis \cite{Howell2001Principles}, the following equations hold
\begin{equation}\label{Pf:ASpecialSolution1}
\sum\limits_{i = 1}^N {{{\cos }^2}\left( {\frac{{2\pi }}{N}i} \right) = } \sum\limits_{i = 1}^N {{{\sin }^2}\left( {\frac{{2\pi }}{N}i} \right) = } \frac{N}{2}
\end{equation}
\begin{equation}\label{Pf:ASpecialSolution2}
\begin{aligned}
&\sum\limits_{i = 1}^N {\cos \left( {\frac{{2\pi }}{N}i} \right)} \sin \left( {\frac{{2\pi }}{N}i} \right) = \sum\limits_{i = 1}^N {\cos \left( {\frac{{2\pi }}{N}i} \right)}  \\
& \ \ \ \ \ \ \ \ \ \ \ \ \ \ \ \ \ \ \ \ \ \ \ \ \ \ \ \ \ \ \ \ = \sum\limits_{i = 1}^N {\sin \left( {\frac{{2\pi }}{N}i} \right) = 0}.
\end{aligned}
\end{equation}
Substituting $\left| {{\alpha _1}} \right|  =  \cdots  = \left| {{\alpha _i}} \right| =  \cdots  = \left| {{\alpha _N}} \right|=\alpha$, $\left| {{\theta _1}} \right|  =  \cdots  = \left| {{\theta _i}} \right| =  \cdots  = \left| {{\theta _N}} \right|=\theta$, ${d_1}  = ...={d_i}=... = {d_N}=d$, (\ref{Pf:ASpecialSolution1}), and (\ref{Pf:ASpecialSolution2}) into (\ref{Eq:FIMofToFandSSP}), we could obtain the same equations as C1--C9.
\end{IEEEproof}

\begin{figure}[t]
\centering \leavevmode \epsfxsize=3.5 in  \epsfbox{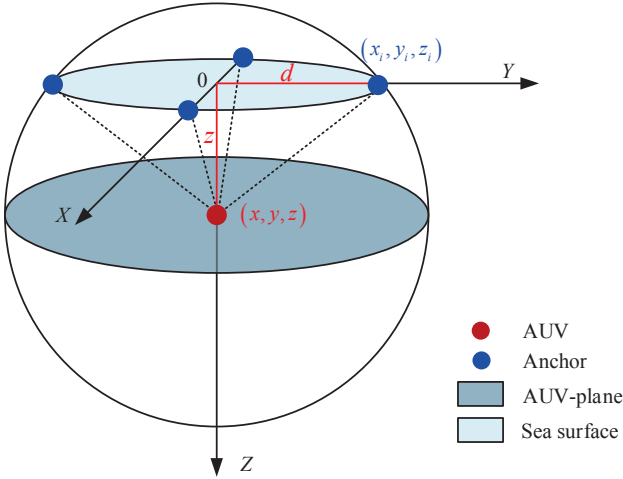}
\centering \caption{The anchor-AUV geometrical shape proposed in Theorem~\ref{Theorem:ASpecialSolution}.}
\label{Fig:Sea_Surface_Circle_Deployment}
\end{figure}

\begin{remark}\label{Rm:Remark2}
Interestingly, Theorem~\ref{Theorem:ASpecialSolution} indicates a special anchor-AUV geometrical shape, where all the anchors are uniformly distributed along a circumference on the sea surface with its center being right above the AUV.
We refer this kind of anchor-AUV geometrical shape as uniform sea-surface circumference (USC) deployment. As shown in Fig.~\ref{Fig:Sea_Surface_Circle_Deployment}, we take four anchors as an example to explain the USC's configuration, where the AUV is placed at a depth of $z$ m and four anchors are placed on the surface. From the theoretical analysis presented before, the anchors are placed at
\begin{equation}
\begin{aligned}
&{p_1} = {\left[ {x + d \cdot \cos \left( {\pi /2} \right),~y + d \cdot \sin \left( {\pi /2} \right),~0} \right]^{\rm{T}}} m\\
&{p_2} = {\left[ {x + d \cdot \cos \pi ,~y + d \cdot \sin \pi ,~0} \right]^{\rm{T}}} m\\
&{p_3} = {\left[ {x + d \cdot \cos \left( {3\pi /2} \right),y + d \cdot \sin \left( {3\pi /2} \right),0} \right]^{\rm{T}}} m\\
&{p_4} = {\left[ {x + d \cdot \cos \left( {2\pi } \right),~y + d \cdot \sin \left( {2\pi } \right),~0} \right]^{\rm{T}}} m.
\end{aligned}
\end{equation}
\end{remark}
It is interesting to point out that the USC deployment could not only be directly adopted to construct high-accuracy GPS intelligent buoys (GIB) localization systems, but is also suitable for the cases when autonomous surface vehicles (ASVs) carrying acoustic anchors keep a relative position to the AUV.
Theorem~\ref{Theorem:ASpecialSolution} provides a kind of anchor-AUV geometrical configuration, while the anchor-AUV range still remains to be determined. In what follows, we first reformulate (\ref{Op:Thrid}) into a univariate optimization problem and then find the anchor-AUV range under the USC.

From the USC deployment scheme, all the anchors are placed on the sea surface with $z_i=0$. Then, the relationship between $d$ and $z$ can be expressed as $d=kz$, where $k$ is the introduced scale factor.
Subsequently, by replacing all anchor-AUV horizontal distances and angles in (\ref{Op:Thrid}) with $k$ and $z$, the tr(CRLB) can be rearranged as a function in $k$ for a given $z$, as shown in (\ref{Eq:CRLBEquality}) at the bottom of this page. In this way, getting the anchor-AUV range is equivalent to finding $k$, which can be obtained by solving the following optimization problem

\setcounter{equation}{26}
\begin{equation}\label{Problem:OptimizeRadius}
\begin{array}{l}
\mathop {\min }\limits_k ~~{{\text{tr(CRLB)}}}\\
{\begin{aligned}
\text {s.t.}~ ~~&{\text{C1}:{\left| {{\alpha _1}} \right|  =  \cdots  = \left| {{\alpha _i}} \right| =  \cdots  = \left| {{\alpha _N}} \right|}},~{ \forall i \in \mathcal{C}}\\
&{\text{C2}:{\left| {{\theta _1}} \right|  =  \cdots  = \left| {{\theta  _i}} \right| =  \cdots  = \left| {{\theta  _N}} \right|}},~{ \forall i \in \mathcal{C}}\\
&{\text{C3}:{\varphi _i} = \frac{{2\pi i}}{N}},~{ \forall i \in \mathcal{C}}\\
&{\text{C4}:{{d_1}  = ...={d_i}=... = {d_N}={d}}},~{ \forall i \in \mathcal{C}}\\
%&{\text{C4}:{r_1}  = ...={r_i}=... = {r_N}=r },~{ \forall i \in \mathcal{C}}\\
&{\text{C5}:{\sigma _1}^2 = ... = {\sigma _i}^2 = .... = {\sigma _N}^2={\sigma}^2},~{ \forall i \in \mathcal{C}} \\
&{\text{C6}:{z_i=0}},~{ \forall i \in \mathcal{C}}.
\end{aligned}}
%\end{small}
\end{array}
\end{equation}

By above transformation, the originally five optimization variables in (\ref{Op:Thrid}), i.e., $\alpha_i$, $\theta_i$, $\varphi_i$, and $d_i$, have been converted into a single $k$ in (\ref{Problem:OptimizeRadius}). We adopt a gradient descent method to find the promising $k$ under the USC geometry, the details of which are summarized in Algorithm~\ref{Algorithm:GradientDescentMethod}.
\begin{algorithm}[!t]
\caption{Gradient descent algorithm.}
\begin{algorithmic}[1] \label{Algorithm:GradientDescentMethod}
\STATE \textbf{Initialization} \label{line1}
\begin{itemize}
\item Set an initial $k>0$.
\item Set the iteration precision $\varepsilon > 0$.
\item Set the step size $t\in(0,1)$, the maximum number of iteration $N_{\max}$, and the iteration index $j=1$.
\end{itemize}
\STATE Set $f(k)=\text{tr(CRLB)}$ in~(\ref{Eq:CRLBEquality}). \label{line2}
\STATE \textbf{repeat}\label{line3}
\STATE \quad 1.~$\Delta k :=-\nabla f(k)$ \label{line4}
%\STATE \quad 2.~Line search.~Choose step size $t$ via exact or backtracking line search.\label{line5}
\STATE \quad 2.~Update.~$k:=k+t\Delta k$ and $j=j+1$. \label{line6}
\STATE \textbf{until} $j>N_{\max}$ or $|\nabla f(k)| \leq \varepsilon$. \label{line7}
\STATE Output the optimal $k$.\label{line8}
\end{algorithmic}
\end{algorithm}

\begin{remark}\label{Rm:AnchorsNumberAndNoiseVariance}
From Eq.~(\ref{Eq:CRLBEquality}), the estimation performance can be improved by increasing the number of anchors $N$ and decreasing the ToF measurements noise variance $\sigma^2$ (because by these the $\text{tr(CRLB)}$ becomes smaller). On the other hand, the proposed USC scheme is easy-to-implement in practice. Moreover, the depth $z$ can be reliably measured by depth sensors nowadays with down to few millimeters' accuracy \cite{3DUnderwaterLocalizationSchemeUsingEMWaveAttenuationWithADepthSensor/ICRA2016}. Based on this fact, we can use the method shown in Algorithm~\ref{Algorithm:GradientDescentMethod} to figure out the USC's radius, i.e., $kz$, for localization accuracy assurance.
\end{remark}

\section{Simulation Examples}
In this section, we present extensive simulation results to verify the performance of our derived theoretical conclusions and proposed anchor-AUV geometry.
We consider the USC deployment (see Fig.~\ref{Fig:Sea_Surface_Circle_Deployment}) for localization, where the anchors' number $N\geq4$, the surface sound speed is $b=1480~\rm{m/s}$, and the underwater sound speed increases linearly with the depth at a steepness of $a$.
From \cite{TDOABasedSourceLocalizationWithDistanceDependentNoises/TWC2015}, fixing the transmit power and operating frequency, the ToF measurement noise is only related to the distance between the anchor and AUV, given by
%the depth measurement noise variance of the AUV is fixed, i.e., ${\gamma_z^2}=1$, while
\begin{equation}\label{Eq:ToF_Measurement_Noise}
{\sigma _i^2}=K_E\cdot A(l_i,f),  ~~ \forall i \in \mathcal{C}.
\end{equation}
\par\noindent In Eq.~(\ref{Eq:ToF_Measurement_Noise}), $K_E$ is a constant that is related to the transmit power and the environment noise floor, and $A(l_i,f)$ denotes the overall path loss of an acoustic signal over a distance at frequency $f$, which is defined as \cite{UnderwaterPropagationModels/StatisticalCharacterization/CommunMag2009}
\begin{equation}\label{Eq:PathLoss}
%\begin{small}
A\left( {{l_i},f} \right) = {\left( {\frac{{{l_i}}}{{{l_0}}}} \right)^\beta }L{\left( f \right)^{{l_i} - {l_0}}}
%\end{small}
\end{equation}
where $l_i$ is the traveled distance that is taken in reference to some $l_0$, $\beta$ denotes the path loss exponent that models the acoustic spreading geometry, commonly in the range of $[1,2]$, and $L(f)$ represents the absorption coefficient that can be obtained by an empirical formula in \cite{ToRelayOrNotToRelayOpenDistanceAndOptimalDeploymentForLinearUnderwaterAcousticNetworks/TCOM2018,UnderwaterPropagationModels/StatisticalCharacterization/CommunMag2009}.
The corresponding parameters used throughout the simulation results are listed as follows: $l_0=1000~\rm{m}$, $\beta=2$, $K_E=-10~\rm{dB}$, and $L(f)=1~\rm{dB/km}$ which is valid for frequencies below $20~\rm{kHz}$. In order to fairly compare the performance of our proposed scheme with existing deployment schemes, in which depth measurement are typically required, we consider that the depth of the AUV can be measured by the equipped pressure sensors and is usually corrupted by the Gaussian noise with zero mean and unit variance\footnote{The assumption that the error introduced by the depth sensor is Gaussian noise may be inappropriate in some practical environments but the corresponding analysis has significantly theoretical value as the reference to what can be achieved and thus has been widely adopted in existing works, e.g., in \cite{TargetLocalization/ToF/SSP/TSP2013,NavigationAndControlSystemOfADeepSeaUnmannedUnderwaterVehicleHEMIRE/OCEAN2006,RealTimeUnderwater3DSceneReconstructionUsingCommercialDepthSensor/USYS2016,AcousticModelsAndSonarSystems/JOE1993}. Investigation based on more realistic noise models, such as the full scale error which is a combination of systematic bias and
stochastic errors, is also very meaningful.}. In addition, an extended Kalman filter (EKF) estimator involving the isogradient SSP is adopted to perform the localization estimation in our simulations.

\begin{figure}[t]
\centering \leavevmode \epsfxsize=3.5 in  \epsfbox{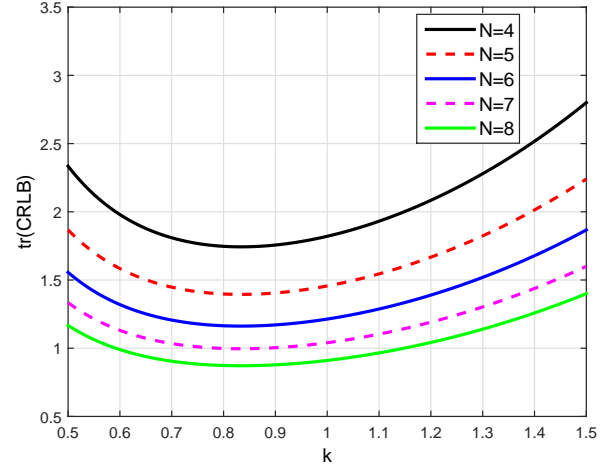}
\centering \caption{The relationship between $tr\left( {{\rm{CRLB}}} \right)$ and $k$ under different anchor number settings. In this figure, the actual position of the AUV is set to be $[50,~50,~50]^{\rm T}\rm{m}$ and $a=0.1$, considering distance-dependent noise.}
\label{Fig:tr(CRLB)_with_k}
\end{figure}
We first show how the parameter $k$ affects the localization accuracy in terms of the tr(CRLB) under different anchor number settings in Fig.~\ref{Fig:tr(CRLB)_with_k}.
As can be seen, for a given $k$, the tr(CRLB) decreases when the anchor number increases, which verifies our theoretical analysis in Remark 3.
It is worth noting that there exists an optimal $k$ that achieves the minimum tr(CRLB) for all of the five curves when $k$ changes from 0.5 to 1.5. For example, the optimal $k$ lies in 0.84--0.85 when the number of deployed surface anchors varies from 5 to 8. Thus, we can use a gradient descent method shown in Algorithm~\ref{Algorithm:GradientDescentMethod} to find $k$ as well as the anchor-AUV ranges.
\begin{figure}[t]
  \centering
  \subfigure[]{
    \label{Fig:RMSE_for_three_deployments_1} %% label for first subfigure
    \centering \leavevmode \epsfxsize=3.5in  \epsfbox{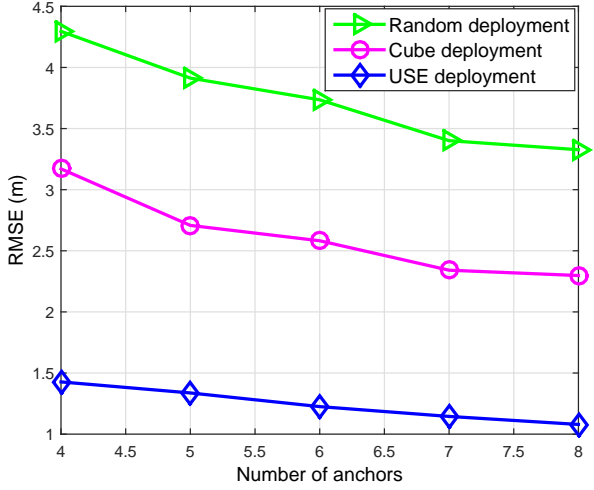}}
  \hspace{0in}
  \subfigure[]{
    \label{Fig:RMSE_for_three_deployments_2} %% label for second subfigure
    \centering \leavevmode \epsfxsize=3.5in  \epsfbox{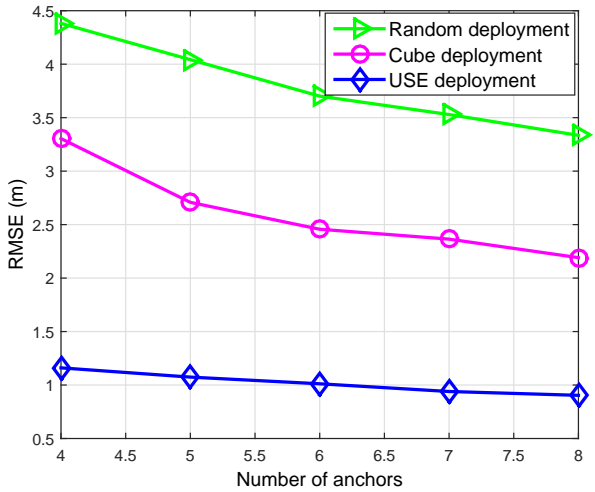}}
  \caption{Comparison of RMSE versus number of anchors among three different schemes. The steepness of SSP is $a=0.1$, considering distance-dependent noise. Regarding the cube deployment and random deployment, we use the typical parameter settings provided in \cite{TargetLocalization/ToF/SSP/TSP2013}. (a) The actual position of the AUV is set to be $[50, 50, 50]^{\rm T} \rm{m}$. (b) The actual position of the AUV is set to be $[100, 100, 100]^{\rm T} \rm{m}$.}
  \label{Fig:RMSE_for_three_deployments}
\end{figure}
%\begin{figure}[t]
%\centering \leavevmode \epsfxsize=3.5 in  \epsfbox{RMSE_for_three_deployments}
%\centering \caption{Comparison of RMSE versus number of anchors among three different schemes. In this figure, the actual position of the AUV is set to be $[50,~50,~50]^{\rm T}\rm{m}$ and $a=0.1$, considering distance-dependent noise. Regarding the cube deployment and random deployment, we use the typical parameter settings provided
%in \cite{TargetLocalization/ToF/SSP/TSP2013} and \cite{ImpactsOfDeploymentStrategiesOnLocalizationPerformanceInUnderwaterAcousticSensorNetworks/TIE2015},  respectively.}
%\label{Fig:RMSE_for_three_deployments}
%\end{figure}

We then evaluate the localization performance of our proposed USC deployment scheme by comparing its localization accuracy with those of the other two widely-adopted anchor deployment schemes, i.e., the cube deployment used in \cite{TargetLocalization/ToF/SSP/TSP2013} and the random deployment. To quantitatively characterize the localization accuracy, we compute the root mean squared error (RMSE) between the actual and estimated positions according to the following formula
\begin{equation}
\text{RMSE} = \sqrt {\mathbb{E}\{ {{{\left\| {\hat {\textbf q} - {\textbf q}} \right\|}^2}} \}}.
\end{equation}
The comparison of the RMSE versus the number of anchors among three different anchor deployment schemes is plotted in Fig.~\ref{Fig:RMSE_for_three_deployments}. From Fig.~\ref{Fig:RMSE_for_three_deployments}, although all the RMSE curves of three deployment schemes decrease with the number of anchors, the USC scheme always has the smallest localization error under different AUV's positions. More importantly, it is observed from the Fig.~\ref{Fig:RMSE_for_three_deployments}(a) that, compared with the cube and the random deployment schemes, the USC can reduce the RMSE by about 50\% and 66\%, respectively.

\begin{figure}[t]
\centering \leavevmode \epsfxsize=3.5 in  \epsfbox{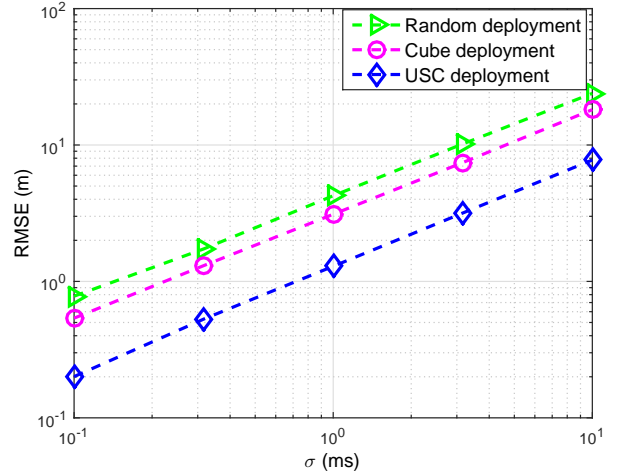}
\centering \caption{Impacts of the time measurement error on the localization accuracy under different deployment schemes. In this figure, the actual position of the AUV is set to be $[50,~50,~50]^{\rm T}\rm{m}$ and $a=0.1$, considering distance-independent noise.}
\label{Fig:Time_measurement_error}
\end{figure}

\begin{figure}[t]
\centering \leavevmode \epsfxsize=3.5 in  \epsfbox{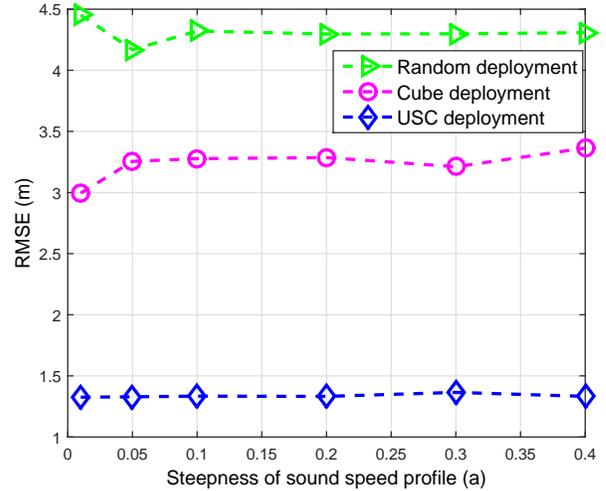}
\centering \caption{Impacts of the steepness of the SSP (a) on the localization accuracy for different deployment schemes. In this figure, the actual position of the AUV is set to be $[50,~50,~50]^{\rm T}\rm{m}$, considering distance-independent noise with standard deviation $\sigma=1~\rm{ms}$.}
\label{Fig:Stepness_of_sound_speed_profile}
\end{figure}

Furthermore, we exhibit the impacts of involved environment parameters on the localization accuracy of the USC scheme, as shown in Fig.~\ref{Fig:Time_measurement_error} and Fig.~\ref{Fig:Stepness_of_sound_speed_profile}. Specifically, Fig.~\ref{Fig:Time_measurement_error} displays how the RMSEs of three deployment schemes change with the variation of the ToF measurement noise standard derivation $\sigma$.
Here, we consider a distance-independent noise variance while maintaining the same distance between the AUV and anchors for these three schemes.
It can be observed that, although the RMSEs of three deployment schemes linearly increase with $\sigma$, the USC deployment scheme always has the smallest localization error.
Meanwhile, even with a large measurement noise, our proposed USC scheme achieves a better localization accuracy than the other two schemes. For example, when $\sigma = 10$ ms, the RMSE is approximately equal to 8 m for the USC but surges to about 20 m for the cube and random deployment schemes. Numerically, it is seen that, when $\sigma$ varies from 0.1 ms to 10 ms, the USC is capable of decreasing the RMSE by 57\%-63\% and 68\%-74\% against the cube and random deployment schemes, respectively.
Fig.~\ref{Fig:Stepness_of_sound_speed_profile} shows the impacts of the steepness of SSP, i.e., $a$ in (\ref{Eq:SoundSpeedProfile}), on the localization accuracy of different deployment schemes. From the figure, the RMSE curve of the USC undergoes the slightest fluctuation with the SSP steepness and always achieves the lowest RMSE in the steepness range of our interest. These results indicate that our proposed USC scheme can guarantee a satisfactory localization accuracy even under dynamically changing underwater environments.

\begin{figure}[t]
\centering \leavevmode \epsfxsize=3.5 in  \epsfbox{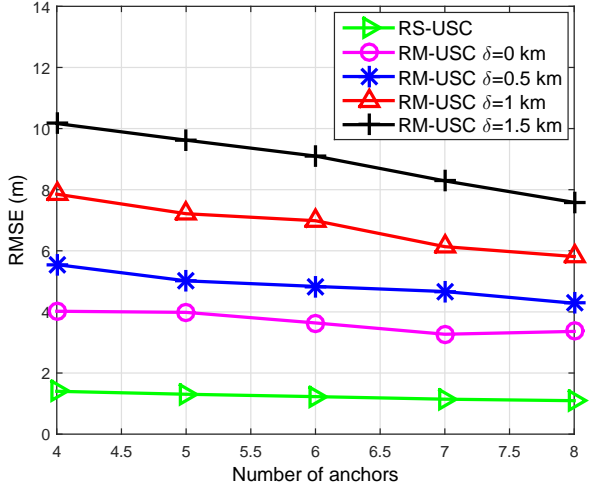}
\centering \caption{Impacts of the distance of the AUV from the anchors' center of gravity on the localization accuracy. In this figure, the initial position of the AUV is set to be $[50,~50,~50]^{\rm T}\rm{m}$ and $a=0.1$, considering distance-independent noise with standard deviation $\sigma=1~\rm{ms}$.}
\label{Fig:AUV_distance_from_the_anchor_center}
\end{figure}
It is noteworthy that the aforementioned analysis explores the anchor-AUV configuration problems under the assumption that the relative positions between the AUV and anchors are fixed.
In practice, the anchors are deployed in advance while the AUV proceeds task with autonomous movement, which results in the variation of the relative position.
In addition, the AUV possibly starts from an unexpected initial position in the hostile undersea environments.
It's therefore important to evaluate the performance of the USC when the AUV is moving with a deviation in the initial position.
We consider the situation where the anchors are deployed according to the AUV's initial position and keep stationary when the AUV has a random movement.
We refer this situation as the RM-USC and use $\delta$ to denote the deviation between the initial position of the AUV and the center of anchor's gravity.
On the other hand, we define the relative-static USC (RS-USC) where the anchors move with the AUV and change their configuration with the AUV's depth.
Fig.~\ref{Fig:AUV_distance_from_the_anchor_center} presents how the localization accuracy varies with the anchor number in the RM-USC and RS-USC scenarios under different deviation values $\delta$.
As can be seen, the RS-USC always has a lower RMSE compared with the RM-USC. Besides, it is intuitive that the RMSE increases exponentially when the deviation of the RM-USC grows. Particularly, within a certain range of deviations $\delta$ ($<1~\rm{km}$), the RMSE of RM-USC is still maintained at a small value ($<6~\rm{m}$), which indicates that the USC is robust to position variance of the AUV and anchors.
\section{Conclusions}
The anchor-AUV geometries, including both the anchor-AUV geometrical shapes and ranges, have significant impacts on the AUV localization accuracy.
Previous works have explored anchor deployment problems for underwater target localization based on the assumption that the sound speed is constant, which actually varies with the depth.
This fact implies that the localization accuracy possibly can be further improved by an appropriate anchor-AUV geometry that takes the underwater varying sound speed into account.
Following this insight, we have devoted to formulating the practical variability of underwater sound speed into the anchor-AUV geometries optimization problem to design a proper anchors' configuration for achieving higher localization accuracy in 3-D underwater environments.
Regarding this problem, we have first derived the Jacobian matrix of the measurement errors at the AUV's true position and used it to quantify the CRLB with ToF measurements under a depth-dependent isogradient SSP.
By minimizing the trace of the CRLB, we have formulated an optimization problem that is multivariate and nonlinear to figure out the satisfactory anchor-AUV geometry.
Based on the Courant-Fischer-Weyl min-max principle and the AM-GM inequality, we have transformed this problem equivalently into a more explicit minimization problem.
Furthermore, an easy-to-implement anchor-AUV geometry, referred to as the uniform sea-surface circumference (USC) deployment scheme, has been obtained by us.
In the USC, all the anchors are uniformly distributed along a circumference on the sea surface with its center being right above the AUV.
The radius of this circumference, or equivalently the anchor-AUV range, produces great impacts on the system's overall localization performance. Thus, we reformulate a range optimization problem under the USC and further figure out the promising anchor-AUV ranges that achieve high localization accuracy through gradient descent algorithm.
Extensive simulation results have verified that the USC has better localization performance than the cube and the random deployment schemes under the same parameter settings.
Moreover, even if the relative geometry of the AUV and anchors changes, the USC deployment can still maintain satisfactory localization accuracy.

It is worthwhile to note that this paper considers the isogradient SSP for simplifying our formulation and analysis. However, practical SSPs possibly vary nonlinearly and even non-monotonically with respect to the depth in some ocean regions, which indicates that the isogradient SSP shown in Eq.~(\ref{Eq:SoundSpeedProfile}) should be modified in these cases. Thus, it is an interesting research direction to extend our proposed methods and analysis to more general SSP models, such as the multiple isogradient SSP and experiment-based SSP. Moreover, when considering more complicated scenarios, such as the scenario that the AUV is moving inside a bounded area, how to improve the AUV localization accuracy from the perspective of anchor deployment is also well worth studying.
\begin{appendices}
\section{Proof of Theorem~\ref{Theorem:JacobianMatrix}}\label{Appendix:ProofOfJacobianMatrix}
In our considered 3-D ToF-based AUV localization scenario, including $N$ anchors, the Jacobian ${\bf{\emph{J}}_o}$ evaluated at the AUV's true position ${\textbf q}=[x,y,z]^{\rm{T}}$ can be given by

\begin{equation}\label{Df:Jacobian}
{\bf{\emph{J}}_o}= \left[ {\begin{array}{*{20}{c}}
{\frac{{\partial {t_1}}}{{\partial x}}}&{\frac{{\partial {t_1}}}{{\partial y}}}&{\frac{{\partial {t_1}}}{{\partial z}}}\\
 \vdots & \vdots & \vdots \\
{\frac{{\partial {t_i}}}{{\partial x}}}&{\frac{{\partial {t_i}}}{{\partial y}}}&{\frac{{\partial {t_i}}}{{\partial z}}}\\
 \vdots & \vdots & \vdots \\
{\frac{{\partial {t_N}}}{{\partial x}}}&{\frac{{\partial {t_N}}}{{\partial y}}}&{\frac{{\partial {t_N}}}{{\partial z}}}
\end{array}} \right]
\end{equation}
\par\noindent where ${\frac{{\partial {t_i}}}{{\partial x}}}$, ${\frac{{\partial {t_i}}}{{\partial y}}}$, and ${\frac{{\partial {t_i}}}{{\partial z}}}$ represent the first partial derivatives for the $i$th anchor's RTT at the AUV localization $\textbf q$. From (\ref{Eq:RayTravelingTime}), calculating the partial derivative for $t_i$ is equivalent to deriving the following corresponding entries
\begin{equation}\label{Eq:PartialDervative}
%\begin{small}
\begin{aligned}
&\frac{{\partial t_i}}{{\partial x}} = \frac{{\partial t_i}}{{\partial \theta_i }}\frac{{\partial \theta_i }}{{\partial x}} + \frac{{\partial t_i}}{{\partial \alpha_i }}\frac{{\partial \alpha_i }}{{\partial x}}\\
&\frac{{\partial t_i}}{{\partial y}} = \frac{{\partial t_i}}{{\partial \theta_i }}\frac{{\partial \theta_i }}{{\partial y}} + \frac{{\partial t_i}}{{\partial \alpha_i }}\frac{{\partial \alpha_i }}{{\partial y}}\\
&\frac{{\partial t_i}}{{\partial z}} = \frac{{\partial t_i}}{{\partial \theta_i }}\frac{{\partial \theta_i }}{{\partial z}} + \frac{{\partial t_i}}{{\partial \alpha_i }}\frac{{\partial \alpha_i }}{{\partial z}}.
\end{aligned}
%\end{small}
\end{equation}
For notation simplification, we define $l_i$ as follows
\begin{equation}\label{Df:l_i}
%\begin{small}
{l_i} = \sqrt {{{\left( {x - {x_i}} \right)}^2} + {{\left( {y - {y_i}} \right)}^2} + {{\left( {\frac{{2b}}{a} + z + {z_i}} \right)}^2}} .
%\end{small}
\end{equation}
By substituting $t_i$ in (\ref{Eq:RayTravelingTime}), $\alpha_i$ in (\ref{Df:Alpha_i}), $\theta_i$ in (\ref{Df:Alpha_i}), $\varphi_i$, $d_i$, $r_i$, and $l_i$ in (\ref{Df:l_i}) into (\ref{Eq:PartialDervative}), we can obtain
\begin{equation}
\frac{{\partial t_i}}{{\partial \theta_i }} = \frac{1}{a} \cdot \frac{{2\sin \theta_i \sin \alpha_i }}{{{{\cos }^2}\theta_i  - {{\sin }^2}\alpha_i }}
\end{equation}
\begin{equation}
\frac{{\partial t_i}}{{\partial \alpha_i }} = \frac{1}{a} \cdot \frac{{2\cos \theta_i \cos \alpha_i }}{{{{\cos }^2}\theta_i  - {{\sin }^2}\alpha_i }}
\end{equation}
\begin{equation}
\frac{{\partial \theta_i }}{{\partial x}} =  - \frac{{\left( {x - {x_i}} \right)\left( {z - {z_i}} \right)}}{{r_i^2 \cdot {d_i}}} =  - \frac{1}{{{r_i}}}\sin {\theta _i}\cos {\varphi _i}
\end{equation}
\begin{equation}
\frac{{\partial \theta_i }}{{\partial y}} =  - \frac{{\left( {y - {y_i}} \right)\left( {z - {z_i}} \right)}}{{r_i^2 \cdot {d_i}}} =  - \frac{1}{{{r_i}}}\sin {\theta _i}\sin {\varphi _i}
\end{equation}
\begin{equation}
\frac{{\partial \theta_i }}{{\partial z}} = \frac{{{d_i}}}{{r_i^2}} = \frac{1}{{{r_i}}}\cos {\theta _i}
\end{equation}
\begin{equation}
\frac{{\partial \alpha_i }}{{\partial x}} = \frac{{\left( {\frac{{2b}}{a} + z + {z_i}} \right)\left( {x - {x_i}} \right)}}{{l_i^2 \cdot {d_i}}} = \frac{1}{{{l_i}}}\cos {\alpha _i}\cos {\varphi _i}
\end{equation}
\begin{equation}
\frac{{\partial \alpha_i }}{{\partial y}} = \frac{{\left( {\frac{{2b}}{a} + z + {z_i}} \right)\left( {y - {y_i}} \right)}}{{l_i^2 \cdot {d_i}}} = \frac{1}{{{l_i}}}\cos {\alpha _i}\sin {\varphi _i}
\end{equation}
\begin{equation}
\frac{{\partial \alpha_i }}{{\partial z}} =  - \frac{1}{{{l_i}}}\sin {\alpha _i}.
\end{equation}
Thus, we can rewrite the Jacobian matrix as (\ref{Df:JacobianMatrix}).

\end{appendices}

\bibliographystyle{IEEEtran}
\bibliography{IEEEabrv,TWC_Paper_Reference}

\section*{Biographies}
\vspace{-2em}
\begin{IEEEbiography}[{\includegraphics[width=1in,height=1.25in,clip,keepaspectratio]{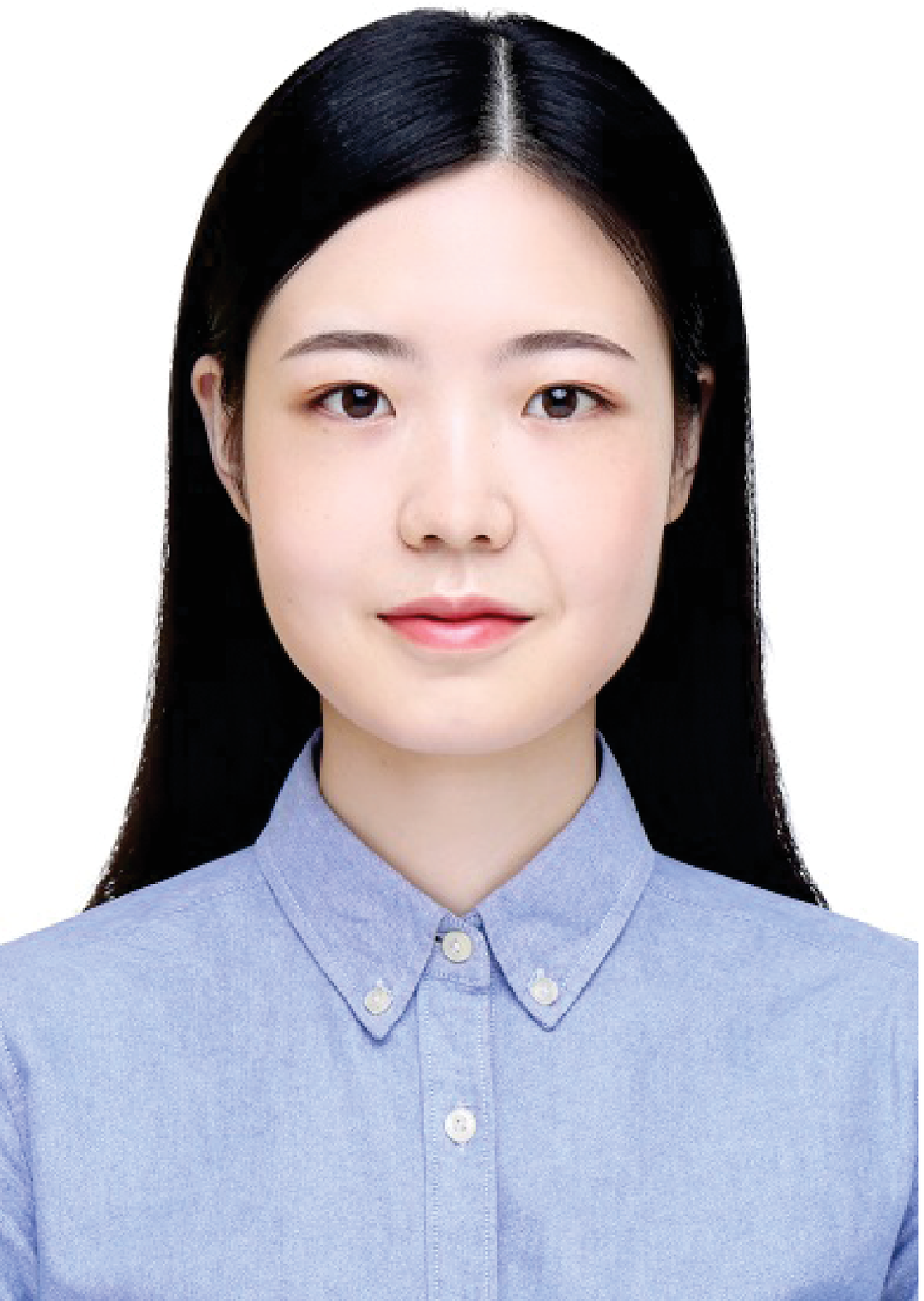}}]{Yixin~Zhang}
received the B.Eng. degree in telecommunications engineering from the School of Tele communications Engineering at Xidian University, Shannxi, China, in 2016. She is currently pursuing the M.S. degree at the School of Electronic Information and Communications, Huazhong University of Science and Technology, Wuhan, China. Her research interests include marine object detection and recognition, and underwater localization.
\end{IEEEbiography}
\vspace{-2em}
\begin{IEEEbiography}[{\includegraphics[width=1in,height=1.25in,clip,keepaspectratio]{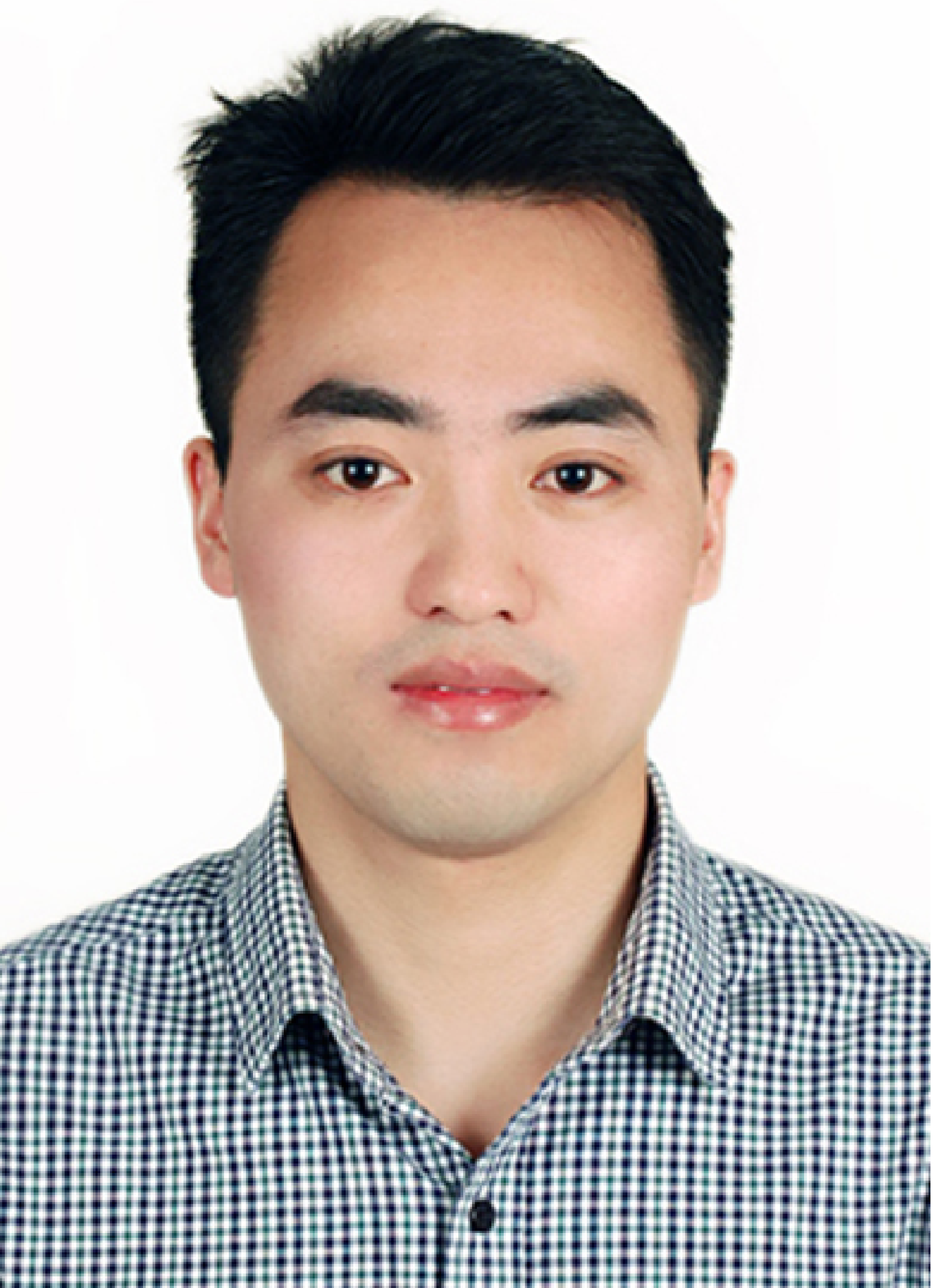}}]{Yuzhou~Li}
(M'14) received the Ph.D. degree in communications and information systems from the School of Telecommunications Engineering, Xidian University, Xi'an, China, in December 2015. Since then, he has been with the School of Electronic Information and Communications, Huazhong University of Science and Technology, Wuhan, China, where he is currently an Assistant Professor. His research interests include machine learning, marine object detection and recognition, and underwater communications.
\end{IEEEbiography}
\vspace{-2em}
\begin{IEEEbiography}[{\includegraphics[width=1in,height=1.25in,clip,keepaspectratio]{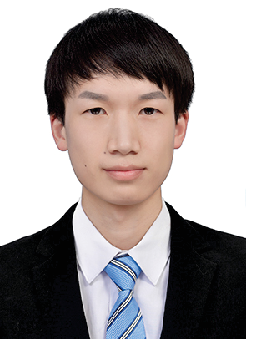}}]{Yu~Zhang}
 received the B.Eng. degree in communications engineering from the School of Electronic Information and Communications, Huazhong University of Science and Technology, Wuhan, China, where he is currently pursuing the M.S. degree. His research interests include marine information networks and underwater communications.
\end{IEEEbiography}
\vspace{-2em}
\begin{IEEEbiography}[{\includegraphics[width=1in,height=1.25in,clip,keepaspectratio]{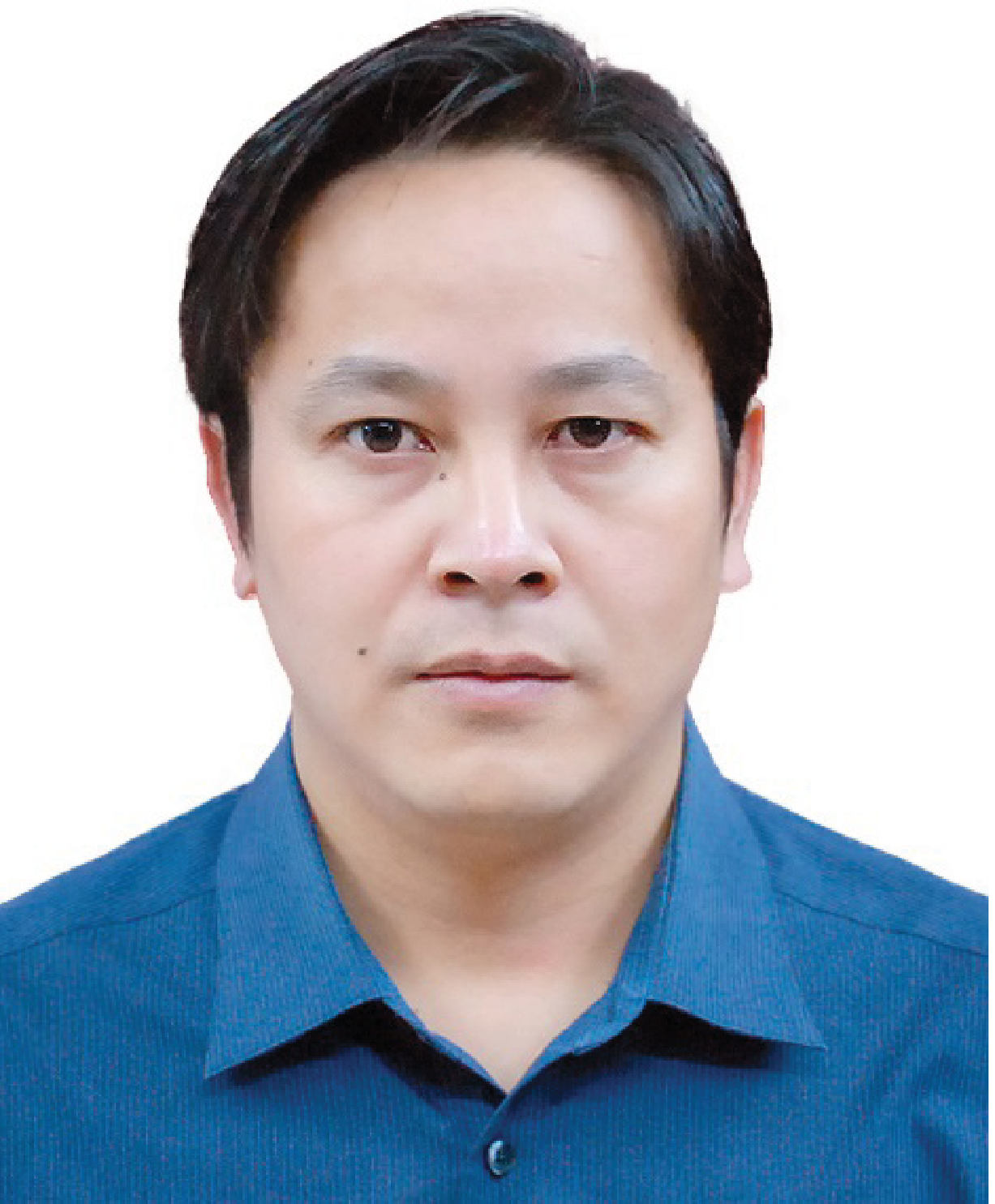}}] {Tao~Jiang}
(M'06--SM'10) is currently a Distinguished Professor with the School of Electronic Information and Communications, Huazhong University of Science and Technology, Wuhan, China. He has authored or co-authored over 200 technical papers and 5 books in the areas of wireless communications and networks. He is the associate editor-in-chief of China Communications and on the Editorial Board of IEEE Transactions on Signal Processing and on Vehicular Technology, among others.
\end{IEEEbiography}

\end{document}